\documentclass{pnastwo}

\usepackage{graphicx}

\usepackage[usenames,dvipsnames]{color}

\usepackage{amssymb,amsfonts,amsmath}

\contributor{Submitted to Proceedings
of the National Academy of Sciences of the United States of America}
\url{www.pnas.org/cgi/doi/10.1073/pnas.0709640104}
\copyrightyear{2011}
\issuedate{Issue Date}
\volume{Volume}
\issuenumber{Issue Number}

\begin{document}



\title{Bending forces plastically deform growing bacterial cell walls}

\author{Ariel Amir
\affil{1}{Department of Physics, Harvard University, Cambridge, Massachusetts 02138, USA},
 Farinaz Babaeipour
 \affil{2}{FAS Center for Systems Biology, Harvard University, 52 Oxford St, Cambridge, MA 02138, USA}
\affil{3}{Department of Physics and Section of Molecular Biology,University of California, San Diego, 9500 Gilman Drive, La Jolla, CA 92093, USA},
 Dustin B. McIntosh \affil{3}{},
 David R. Nelson \affil{1}{}\affil{2}{} , and Suckjoon Jun \affil{2}{}\affil{3}{}{}}

\contributor{Submitted to Proceedings of the National Academy of Sciences
of the United States of America}

\maketitle

\begin{article}

\begin{abstract}

Cell walls define a cell shape in bacteria. They are rigid to resist large internal pressures, but remarkably plastic to adapt to a wide range of external forces and geometric constraints. Currently, it is unknown how bacteria maintain their shape. In this work, we develop experimental and theoretical approaches and show that mechanical stresses regulate bacterial cell-wall growth. By applying a precisely controllable hydrodynamic force to growing rod-shaped \emph{Escherichia coli} and \emph{Bacillus subtilis} cells, we demonstrate that the cells can exhibit two fundamentally different modes of deformation. The cells behave like elastic rods when subjected to transient forces, but deform plastically when significant cell wall synthesis occurs while the force is applied. The deformed cells always recover their shape. The experimental results are in quantitative agreement with the predictions of the theory of dislocation-mediated growth. In particular, we find that a single dimensionless parameter, which depends on a combination of independently measured physical properties of the cell, can describe the cell's responses under various experimental conditions. These findings provide insight into how living cells robustly maintain their shape under varying physical environments.

 \end{abstract}



\noindent\fbox{%
    \parbox{0.45 \textwidth}{%
   \small{Significance Statement:  Regulation of cell wall growth is a process of fundamental importance in cell biology. In this work, we demonstrate for the first time that mechanical stress directly influences cell-wall synthesis of bacteria. In a series of simple experiments, we elastically and plastically deform cell walls as they grow by applying anisotropic mechanical stresses to bacteria.  Using a theory of dislocation-mediated growth, we explain how growth and form of the cell walls are quantitatively related to one another in bacteria.}
    }
}

\vspace{0.3 cm}
Biological systems exhibit many properties rarely found in condensed matter physics, which are often caused by \emph{growth}. When coupled to mechanical forces, growth can drive a wide range of cellular phenomena such as regulation of the eukaryotic cell morphology by actin networks~\cite{maha1}, collective behavior in tissues~\cite{weitz}, cell differentiation~\cite{discher}, and shape and division of yeast and plant cells~\cite{boudaoud, elbaum}. Of fundamental interest as well as practical importance is understanding the relationship between growth, form, and structure of bacterial cell walls~\cite{sun}. Bacterial cell walls define a cell morphology and hold the large internal (turgor) pressure. Many antibiotics target them to efficiently hamper cell growth and reproduction. As such, cell walls and their synthesis have been the subject of extensive biochemical \cite{Lovering2012} and biophysical \cite{sun} studies in the context of cell growth \cite{Scheffers2005}, cell shape \cite{young} and cell division \cite{DenBlaauwen2008}.

Despite the long history \cite{koch}, however, we are still far from being able to predict the shape or dimensions of any cells from first principles based on the information obtained from studies so far. Recent experimental work shed new insights in this regard. For example, bacteria can significantly deform when grown with constraints~\cite{whitesides, dekker} and yet are able to recover their native shape~\cite{dekker}. However, the mechanism underlying deformation and recovery, as well as the cues which regulate cell-wall growth, have not been well understood.

We have developed combined experimental and theoretical methods to directly address how mechanical stresses are involved in regulation of cell wall growth. Our experimental approach is illustrated in Fig.~1. Rod-shaped \emph{Eschericia coli} or \emph{Bacillus subtilis} cells are inserted into snuggly fitting micro-channels where they are grown in a controlled environment.  Filamention is induced in the cells by suppressing division, see the Materials and Methods section for further details.


The filamentous cell is subjected to precisely calibrated hydrodynamic forces. This simple approach has two notable advantages to previous methods: (1) we can directly probe the mechanical properties and responses of the cell walls \emph{non-invasively} for a wide temporal range (from $<$ 1 second to over one hour) (2) we can achieve force scales several orders of magnitude larger than what is possible by optical traps~\cite{wang} or atomic force microscopy~\cite{yao}. Using this approach, we demonstrate that the cells can exhibit two fundamentally different modes of deformations, and recover from them. The first mode is \emph{elastic} in that the cell recovers its shape as soon as the external force is removed. The other mode is \emph{plastic} and requires growth for morphological recovery. The plastic deformations are due to \emph{differential growth}, resulting from a coupling to mechanical stresses varying along the cell walls.

We also provide a theoretical framework that explains our experimental findings quantitatively. The basic idea is that external forces are transduced to the activity of the cell-wall growth machinery. To explain shape deformations, we calculated the effect of force-transduction by extending the theory of dislocation-mediated growth of bacterial cell walls~\cite{amir_nelson_pnas}. We found that, for pulse-like forces that are transient, the cell walls respond elastically in agreement with~\cite{wang}. In contrast, if the duration of applied forces is comparable or longer than the timescale of cell-wall growth, the force-transduction causes differential growth of the cell walls and plastic deformations as seen in~\cite{dekker}.

\begin{figure}[t]
\includegraphics[width=8 cm]{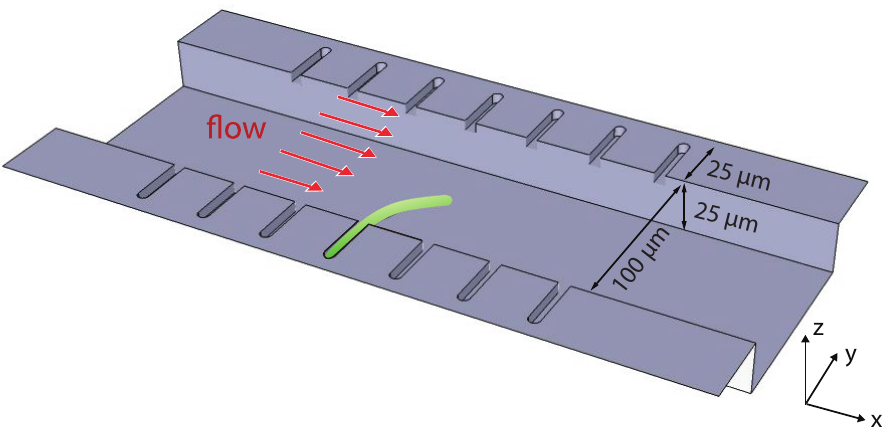}
\caption{Illustration of the experimental approach. Single \emph{E. coli} cells are inserted into micro-channels of length $25$ $\mu$m. The cross section of the micro-channels is square (approximately 1.2 $\mu$m x 1.2 $\mu$m.) and the cells fit snuggly. Cell division can be blocked by expressing SulA proteins that inhibit formation of the (FtsZ) constriction ring, so that we can induce large deformations and recovery of the cell for a long period. The cells were grown typically up to 50 $\mu$m during measurements. Controlled hydrodynamic forces are exerted on the cells by the viscous drag of the growth media (SI text).
\label{fig:device}}
\end{figure}

\section{Results and Discussion}

\subsection {Cells deform elastically by pulse-like forces}

To show that our approach can accurately probe the mechanical properties of \emph{E. coli} and \emph{B. subtilis} cells, we applied short pulses of hydrodynamic force to the cells and monitored their response (Fig.~2). For this, the cells were grown in excess nutrients initially without flow. As the cells formed long, straight filaments, we applied a series of pulse-like flow. Each of these pulses lasted for less than 10 seconds so that the effect of growth is negligible. Fig.~2 and Supporting Video S1 show our typical experimental result. The cells bent transiently at each pulse, and fully recovered their shape immediately after the applied force was removed.

We can explain the experimental results using the theory of linear elasticity. In the theory, the long filamentous cell can be considered as a cylindrical beam bending by a hydrodynamic force (SI text). Shown in Fig.~2B is the comparison between data and theory for the tip position (thus the degree of cell bending) of the cell during a representative experiment in Fig.~2A. From these experiments, we extracted a flexural rigidity of  $\approx 5 \times 10^{-20}$ N$\cdot$m$^2$, comparable to the result of $\approx 3 \times 10^{-20}$ N$\cdot$m$^2 $ obtained by mechanical manipulation of \emph{E. coli} cells using optical traps~\cite{wang}. Assuming a cell wall thickness of $4$ nm, this corresponds to a Young's modulus of $\approx 30$ MPa. It should be noted that the small mismatch between the bacterium's diameter and the width of the end of the channel has to be taken into account, see the SI text for further details. We also show in the SI text that the turgor pressure does not contribute to the restoring torque. Similar experiments were performed for \emph{B. subtilis} (see SI Fig.~5) giving a Young's modulus of $\approx 20$ MPa, consistent with past results \cite{Tuson2012}. From these results, we conclude that both \emph{E. coli} and \emph{B. subtilis} cells respond to fast mechanical perturbations as a linear elastic rod.
\begin{figure*}[b]
\includegraphics[width=17 cm]{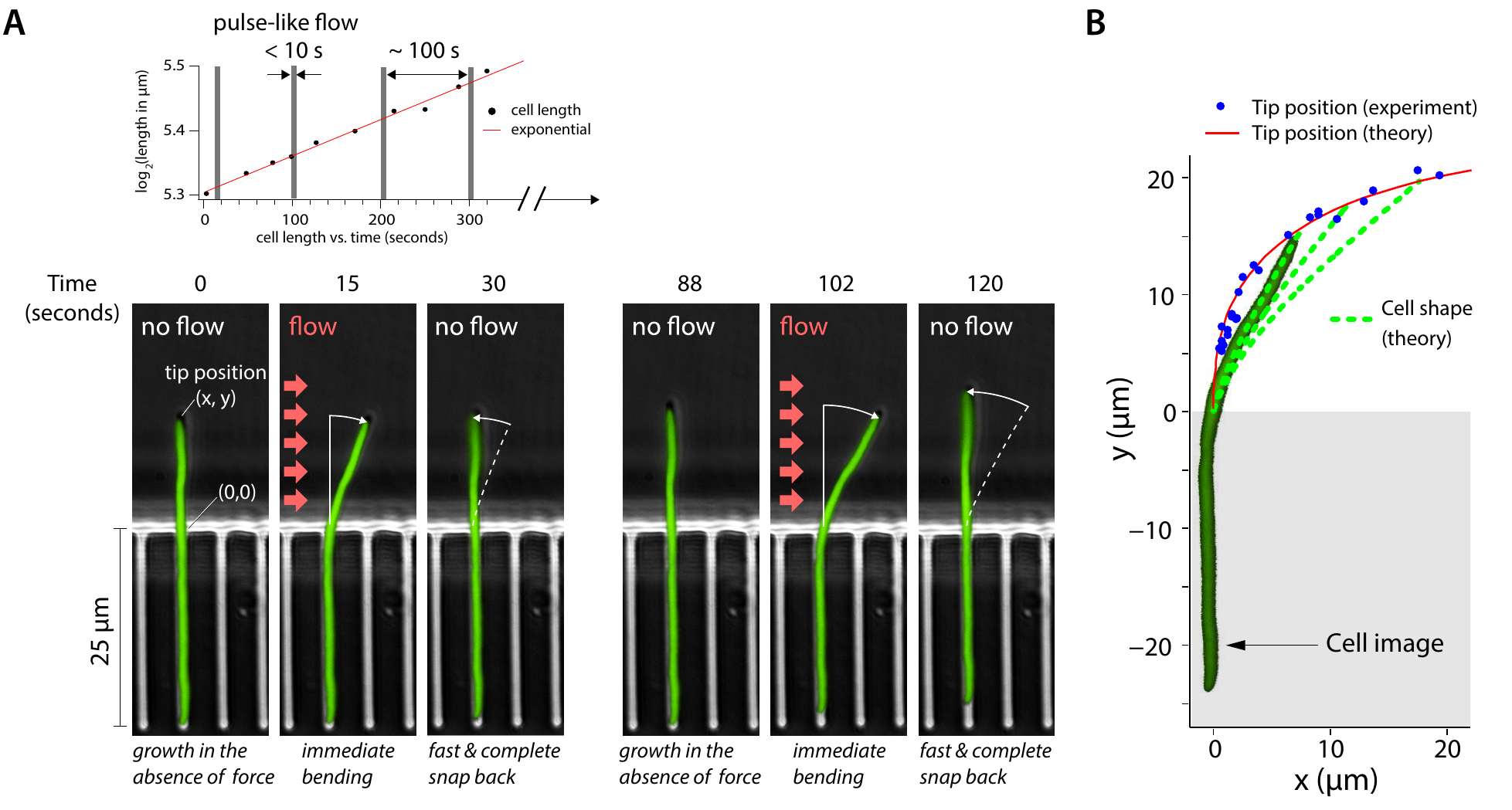}
\caption{\label{fig:elastic}Cells deform elastically by pulse-like forces. (A) Microscopy images of the \emph{E. coli} cell  from Supporting Video S1 taken at different stages of the experiment. The cell initially grows straight out of the microchannel. We applied pulse-like flow to the straight cell repeatedly every several minutes (see top panel). This resulted in completely reversible deformations (two examples are shown). (B) The tip positions (x, y) of the cell from multiple experiments are compared to the theoretical predictions of the linear elasticity theory (SI text). An actual cell shape is superimposed on one of the theoretical curves.}

\end{figure*}

\subsection {Plastic deformations require growth and long-term forces}

The elastic deformation and recovery of the cells to pulse-like forces is consistent with the view that \emph{E. coli} and \emph{B. subtilis} cells can be considered as elastic cylindrical beams~\cite{wang, yao}. The linear elasticity theory, however, cannot explain the observations that bacteria can also deform plastically~\cite{whitesides, dekker}. For plastic deformations, we noticed that the cells in Refs.~\cite{whitesides, dekker} were grown with geometric constraints for a duration comparable to or longer than their mass-doubling time. Since these geometric constraints may exert forces on the cells, we hypothesized that external forces were transduced to the cell-wall synthesis machinery during growth. Therefore, the force-transduction may cause differential growth, thus deformations, of growing cell walls.

As a first step to testing our hypothesis, we grew the cells with a constant hydrodynamic flow (of the growth medium). The filamentous cells grew into a curved shape because of the continuous hydrodynamic forces (Fig.~3). The curved shape of the cell, however, was not easily distinguishable from elastically deformed cells in Fig.~2. To confirm that the cells deformed plastically, we abruptly switched off the flow when the cells significantly bent. The cells did not recover their straight shape immediately, i.e., they deformed plastically (Fig.~3A and Supporting Video S2).

We repeated the experiments with non-growing filamentous cells, and the cells did not show plastic deformations despite long-term application of forces (see SI Fig.~8). We conclude that plastic deformations require both cell wall synthesis and application of forces for a duration over which substantial cell wall synthesis occurs.
\begin{figure*}[t]
\includegraphics[width=17 cm]{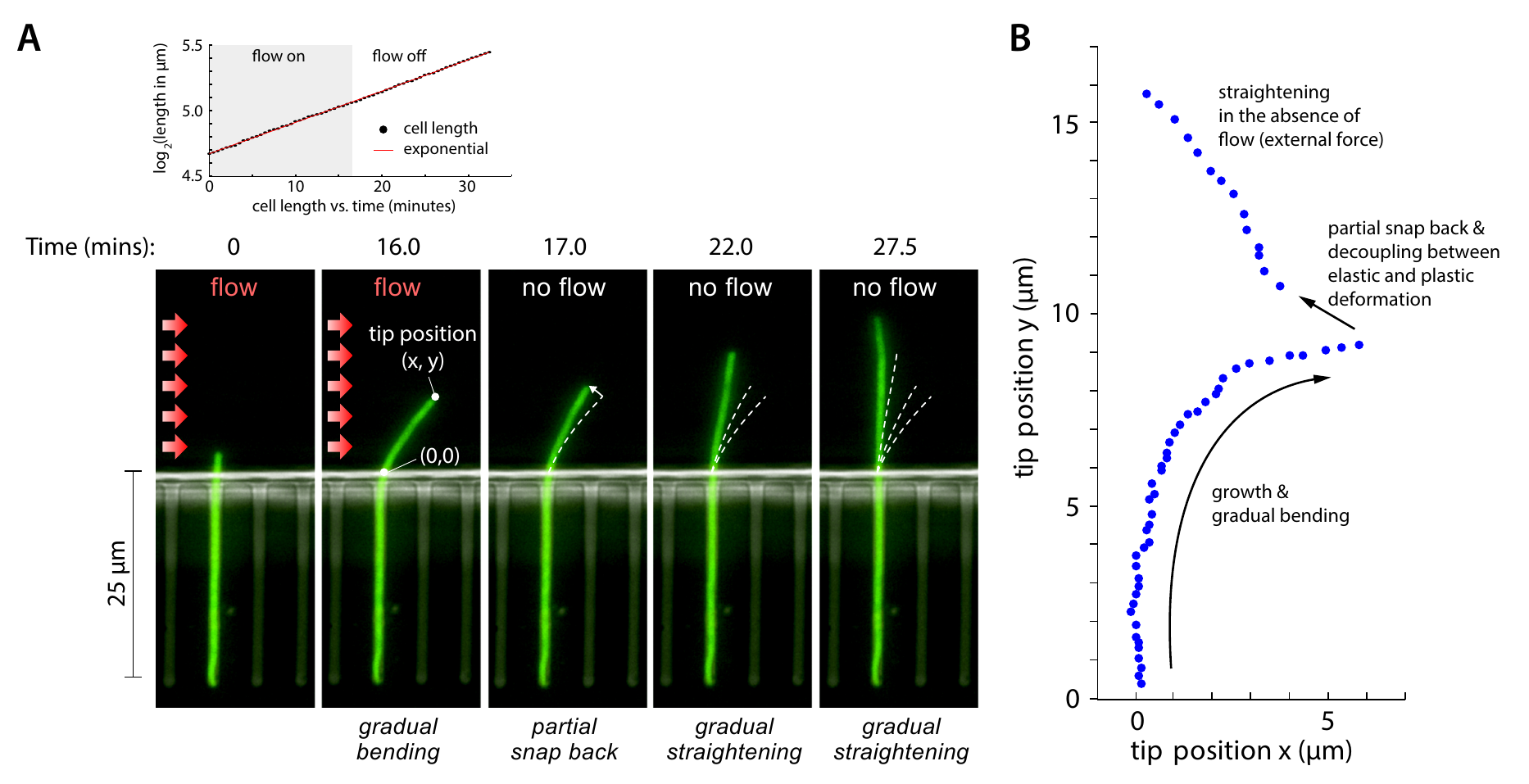}
\caption{\label{fig:plastic}Decoupling the two fundamental modes of deformation. (A) (Supporting Video S2) i. ($t$ = 0 to 16 mins) The cell is growing in the presence of a constant flow for a period comparable to the cell division time. The deformation represents the sum of elastic and plastic modes. ii. ($t$ = 17 mins) As the flow is abruptly turned off, the elastic mode vanishes and the cell \emph{partly} recoils. The residual deformation reflects the cumulative effect of force-transduction, which resulted in differential growth of the cell wall.  iii. ($t >$ 17 mins) In the absence of the flow, the cell straightens gradually as it grows. The top panel shows that the length of the cell increases exponentially despite the large deflections by the external mechanical forces - the mass-doubling time does not change appreciably during the experiment (see top panel) and is comparable to that of non-filamentous cells. (B) Scatter plot of the experimental results for the tip position as time progresses through the snapback experiment from (A).}
\end{figure*}  

\subsection{Elastic and plastic components of deformation can be decoupled}

The plastic deformation experiments in Fig.~3 unraveled an important feature of bacteria cell walls. As the external forces were removed, the cells always partially snapped back. This is clearly seen in the time trace of the cell's tip position (Fig.~3B). The partial snap back suggests that the cell bending in Fig.~3 has \emph{both} elastic and plastic components. In other words, external forces not only cause elastic bending, but may also lead to differential growth which further bends the cell.  The two components may be decoupled by removing the external forces, upon which the elastic component should vanish instantly.

To test our hypothesis, we developed a quantitative, phenomenological model that incorporates force transduction and growth. We describe below our model and its comparison with data. We found the data and theory in good quantitative agreement.

\subsection{Theoretical framework for coupling mechanical stresses to cell wall growth}
\label{subsec:theory}

Our starting point for understanding the data is the theory of dislocation-mediated growth that we developed recently~\cite{amir_nelson_pnas,nelson_review}. The model was inspired by recent experimental observations that, in both Gram-positive~\cite{garner, escobar} and Gram-negative~\cite{shaevitz} bacteria, MreB proteins, associated with cell wall growth, move processively along the cell's circumference (Fig.~4). Processive circumferential insertions resemble the climb of edge dislocations in a two-dimensional crystal \cite{hirth}, and thus one can utilize the theory of dislocations well developed in materials science.

In the previous version of the growth model, we studied the motion of dislocations on a perfect cylinder in the absence of external forces~\cite{amir_nelson_pnas, amir_nelson_PRE}. Therefore, the previous version cannot account for deformations. We extended our theory to allow for differential growth by transduction of external stress to dislocations (growth machinery).

In our theory, tensile (stretching) stress can have two different effects: (1) increase the speed of processive motion of dislocations (Fig.~4A) or (2) reduce the activation energy for nucleation (Fig.~4B). Compressive stress will have the opposite effect. The forces on the elongation machinery are given by the Peach-Koehler force~\cite{peach} acting on dislocations:
\begin{equation}
F_\text{i}=b_{k}\sigma_\text{jk}\epsilon_\text{ijz},\label{eq:peach}
\end{equation}
where $\epsilon_\text{ijz}$ is the Levi-Civita tensor, $\sigma_\text{jk} $ is the stress tensor and $\vec{b}$ the Burgers vector characterizing the dislocation.

For the cylindrical geometry of the cells in our experiment, the theory predicts that the $\sigma_\text{yy}$ component of the stress tensor is the one which couples to the circumferential growth, where $y$ is the coordinate along the cell's long axis, see Fig. 1. As the cell grows in the presence of a constant flow, a cell wall of thickness $h$, length $l$ (measured from the channel end to the cell tip) and radius $r$ experiences a mechanical stress that depends on its position in the cell (Fig.~4A):
\be \sigma_\text{yy}(\theta) = \frac{f \sin(\theta)(y-l)^2}{2\pi r^2 h}, \label{eq:stress}\ee

\noindent where $\theta$ is the azimuthal angle of the cylinder along the circumference ($\theta=\pi/2$ for the cell front facing the flow, and $\theta=-\pi/2$ for the back), $f$ the force per unit length induced by the flow, $y$ the position along the cylinder's axis and $\sigma_\text{yy}$ the mechanical stress induced by the flow. The surface directly facing the incoming flow feels a tensile stress (surface-stretching force), whereas the other side feels a compressive stress (which acts to shorten the distance between adjacent glycan strands). Accordingly, the cell walls grow differentially, growing faster on the upstream side of the flow. In our experiments, the external hydrodynamic force would increase the Peach-Kohler force by $10$ pN for the cell front (at the channel end, for $l=10$ $ \mu$m), and decrease the force by the same amount on the opposite side. This force is comparable, for example, to the one generated by RNA polymerase~\cite{polymerase}. It is also comparable to the force due to the turgor pressure ($\approx 30$ pN for \emph{E. coli} \cite{amir_nelson_pnas}), and thus we expect the mechanical stress to significantly change the activation energy for enzymatic reactions (see \cite{koch}, p. 165).

This differential growth leads to a dynamically deformed cell shape. Due to the large aspect ratio of the filamentous bacteria in our experiments, a relatively small asymmetry in growth leads to a large angular deflection. In the SI text we show that the angular deflection due to differential growth in a small segment of the cell is given by:
\be \Delta\varphi= \frac{\Delta l}{2r}, \label{amplify} \ee

\noindent where $\Delta l \equiv (dl_\text{1}-dl_\text{2})$ is the net difference on opposite sides of the cell wall due to the asymmetric growth. Therefore, for a small segment of length $l$ the relative asymmetry $\Delta l / l$ is ``amplified" by the aspect ratio $l/2r$. This amplification of differential growth often appears in nature, in a diverse range of systems such as the mechanics of plants and of the human gut~\cite{maha3,maha2}.

\subsection{Data supports anisotropic mechanical stress-driven differential growth of cell walls}

A key result that emerged from our theory is a single dimensionless parameter $\chi$ that quantifies the relative contribution of the elastic and the plastic components of deformation. The parameter $\chi$ depends only on the intrinsic physical parameters of the cell, independent of the experimental conditions. We show in the Supplementary Information that
\be \chi = p r / Y h, \label{eq:e_v_p}\ee where $p$ is the turgor pressure, $r$ the cell's radius, $Y$ the Young's modulus, and $h$ the thickness of the cell wall. For $\chi \gg 1$, the elastic effects dominate, and there should be full snapback. For $\chi \ll 1$, the nature of deformation is plastic, and there would be no snapback. We find $\chi$ is of order 1 for the measured values for \emph{E. coli} (see Supplementary Table 1), \i.e., the snapback should be \emph{partial}. \emph{B. subtilis} also has a similar $\chi$ value because \emph{B. subtilis} has a proportionally thicker cell wall and higher turgor pressure (Supplementary Table 1).
\begin{figure}[t]
  \includegraphics[width=8 cm]{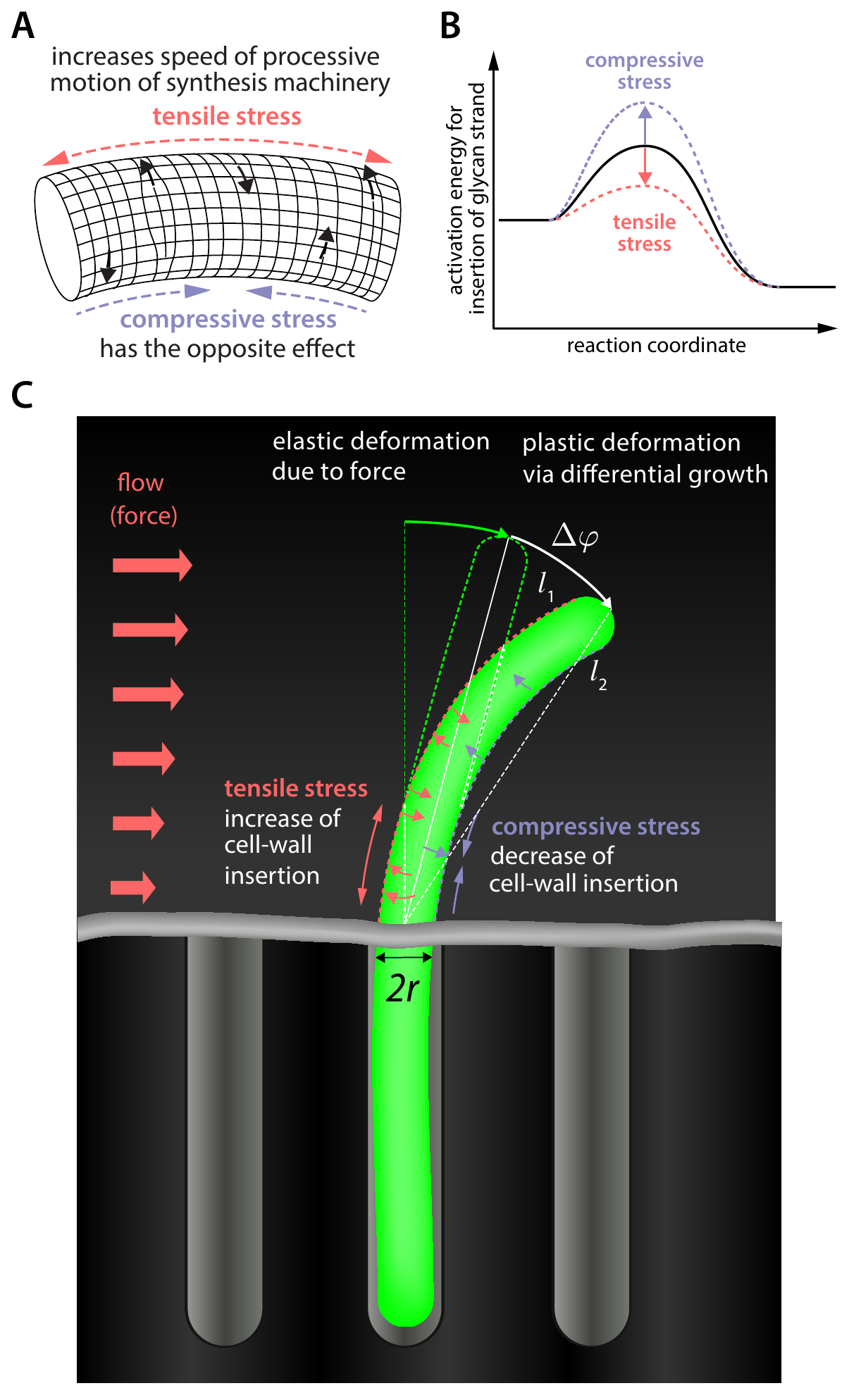}
  \caption{\label{fig:model}Illustration of the differential growth driven by coupling between mechanical stress and cell wall synthesis. (A) Tensile (stretching) stress increases speed of processive motion of strand elongation machinery. Compressive stress has the opposite effect. Black arrows indicate the trajectories of activated dislocations. (B) Similarly, tensile stress will lower the activation energy for insertion of a new glycan strand, and compressive stress will increase it. (C) Asymmetric insertion of dislocations is enhanced by the large aspect ratio of the filamentous bacteria, leading to the large observed angular deflections.}
\end{figure}

Our predictions can be tested quantitatively by measuring the magnitude of snapback and relating it to $\chi$. Note that the local curvature is proportional to the local growth asymmetry on the front and back sides of the cell. We find that the slope of the line connecting the tip and the hinge of the cell is a good measure for the integrated differential growth [$\tan(\phi)$, see Fig.~4]. With this, our theory (SI text) predicts that for small deflections the angles before and after snapback should be proportional, with a slope depending on $\chi$: large values of $\chi$ imply a small slope (since the relative importance of plastic deformations is smaller for larger turgor pressure). Fig.~5 confirms this prediction. It shows the snapback events of 33 \emph{E. coli} and 33 \emph{B. subtilis} cells performed under 8 different experimental conditions. As predicted, data for each species collapse onto a master line regardless of the growth medium, temperature, and Stokes force. The slope of the master line is approximately $1/2$, consistent with a value of $\chi$ of order unity.

\subsection{Implications on cell-shape regulation}

That our prediction is verified by two species of bacteria as evolutionarily distinct as \emph{E. coli} and \emph{B. subtilis} indicates that force transduction may have more general biological consequences for growth.  For example, recent studies indicate that mechanical force plays a major role in cell-shape deformation leading to tissue morphogenesis in higher organisms~\cite{morphogenesis}.

If mechanical stress regulates the rate of cell wall synthesis, it should play a central role in shape regulation. On a cylinder the circumferential stress due to turgor pressure is twice as large as the stress along the cell's long axis. This difference in stresses will provide a natural cue for orienting the growth machinery and coordinating glycan strand insertions. In Gram-positive bacteria, the stress regulation will further lead to a constant thickness of the cell wall by a negative feedback mechanism: the mechanical stress is inversely proportional to the thickness of the cell wall, and thinner parts of the cell wall will have larger stresses due to turgor pressure, leading to enhanced local growth.

\begin{figure}[bt]
  \centering
\includegraphics[width=8.5 cm]{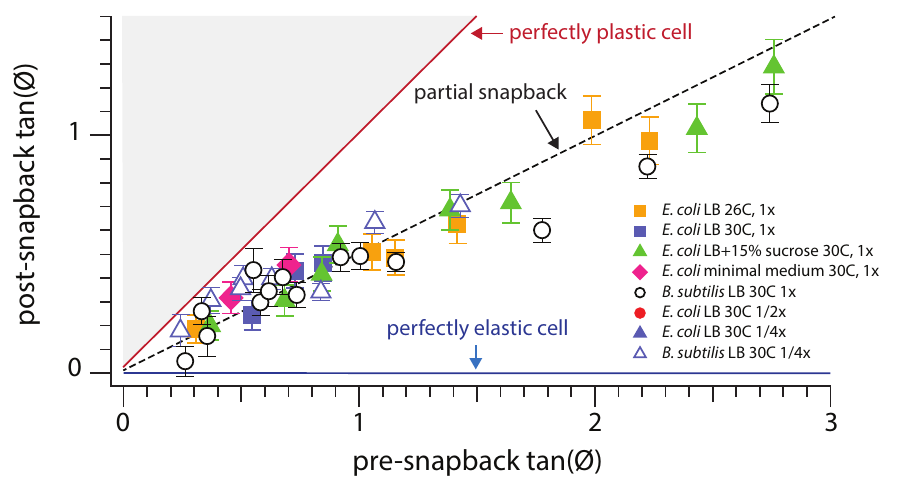}
\caption{Relative importance of elastic versus plastic modes. The ``snapback" of the cell upon switching off the flow (see Fig.~3), parameterized in terms of the slope of the tip position, as measured for 66 cells (33 \emph{E. coli}, 33 \emph{B. subtilis}) in differing conditions growth conditions (see legend for growth media, temperature, and flow rate - 1x being the maximal flow rate). For each snapback, we quantified its magnitude by measuring the deflection angle $\phi$ of the tip before and after the snapback. Error bars are omitted from several series for clarity, but are on the order of those shown. The snapback angle is approximately proportional to the initial angle of deflection when the angles are small, confirming the prediction of the dislocation-mediated growth theory (SI text). The $\approx$1/2 slope shows that the contributions by the two fundamental modes of deformation (elastic vs. plastic) are comparable with one another, regardless of the temperatures and growth conditions.\label{fig:snapback}}
\end{figure}

Note that, throughout the experiments, there were cell-to-cell variations in the bending rates, which was about $30-50$ \% of the mean. This suggests that the variability of elastic constants between different cells might be smaller than what was found in earlier studies~\cite{wang}.

\section{Conclusion}

We have developed a simple, novel experimental approach to directly probe the mechanical properties of bacterial cell walls. Our results indicate that the cell-wall growth rate depends on the mechanical stress, where faster growth is obtained for tensile (stretching) stress. This mechanism is likely involved in shape regulation and the maintenance of a stable rod-shaped cell over time. We found that bending forces cause the cell walls to grow differentially. The differential growth leads to persisting plastic deformations, whose recovery requires cell wall synthesis. This is in sharp contrast to the elastic deformations by pulse-like forces.

The elastic and plastic deformations represent two fundamental modes of cellular response to external mechanical forces, and can be decoupled experimentally as predicted by our theory. These effects can be revealed only by applying bending forces, and cannot be studied by means which can only change the uniform pressures that have to exist inside and outside the cell walls (e.g., osmotic shocks): since cell-wall growth is not the bottle-neck for the cell growth rate \cite{hwa}, the growth rate is not dictated by the cell-wall synthesis rate.

In the future, it would be illuminating to study deformation behavior in other types of bacteria such as tip-growers~\cite{brun}. Based on the mechanism described here, unlike \emph{E. coli} or \emph{B. subtilis}, we expect purely elastic deformations regardless of the duration of applied external forces. At the molecular level, direct observation of the asymmetric growth would provide critical information for the mechanism of cell wall synthesis. In particular, any changes in the speed of cell-wall synthesis rate or nucleation rate would be a direct test of our model. This would be a challenging, but experimentally feasible based on the technology available to us~\cite{garner, escobar, shaevitz}. Finally, it would be fascinating to study similar mechanisms in higher organisms. What is the role of forces in guiding the vast diversity of organism shapes in nature? Previous works have shown the importance of mechanical forces in controlling the shapes of mammalian cells, where the properties of the extracellular matrix can have a large effect on the shape of the cells~\cite{ingber1}, their fate in differentiation~\cite{discher}, and tissue morphogenesis~\cite{morphogenesis}.

The bacterial model system studied here provides insights into the growth process itself, indicating that coupling of mechanical forces should be taken into account to understand robust morphological regulation of single-cell organisms. \\



\begin{materials}

\subsection {Experimental setup}
To deform growing cells non-invasively by applying a well-defined force, we employed a microfluidic device, ``mother machine," previously developed by some of us~\cite{jun}. This device ensures balanced growth of individual cells for hundreds of generations. Furthermore, by controlling the flow speed in the device, the ambient buffer or the content of growth medium surrounding the cells can be switched within fractions of a second~\cite{pelletier}. Thus, we can tune the force applied to the growing cell since the force is directly proportional to the speed of the flow and the size of the cell \cite{berg}. The force which we achieve in this setup far exceeds that attainable with AFM and optical tweezers. Also, our technique provides much more control of the applied forces in comparison, for example, with cells confined to microfluidic chambers of fixed or deformable shape.

For the purpose of our experiments, it was beneficial to work with filamentous cells with cell division suppressed. We grew the cells typically up to 50 $\mu$m long. This allowed us to precisely monitor the deformation of the cells in the presence of external mechanical forces.  For the \emph{E. coli} experiments, we used strain MG1655 with sulA under an inducible promoter (gift from Debu Raychaudhuri, see a description of the plasmid pDB192 in Ref. \cite{deboer}) and cytoplasmic YFP. SulA inhibits polymerization of FtsZ, and upon induction with 1 $mM$ concentration of IPTG the cells became filamentous. YFP is constitutively expressed in order to increase the signal-to-noise ratio of the images taken. For the \emph{B. subtilis} experiments, we used a YlaO (homolog of FtsW) knockout with YlaO reinserted under a D-xylose inducible promoter (Pxyl) (gift from Michael Elowitz).  This strain is typically grown in the presence of 0.5\% xylose; filamentation is induced by removing xylose from the growth media.

We used the hydrostatic pressure due to the controlled height difference between the inlet and outlet to control the flow through the device and achieve short (subsecond) response times of the flow velocity, as we verified using fluorescent beads. In the SI text we analyze the flow profile through the microfluidic device, both analytically and experimentally. We calculate the resulting forces and torques on the cell due to the viscous flow, and provide details of the microscopy and image processing algorithms used.


\end{materials}


\begin{acknowledgments}

We thank E. Garner, S. T. Lovett, S. Wang, N. S. Wingreen and A. Wright for stimulating discussions. We are also indebted to A. Rotem and H. A. Stone for advice regarding the microfluidics, and to E. Efrati, J. W. Hutchinson and J. Paulose for useful discussions of the mechanical aspects and differential growth.  We thank S. Taheri, C. Nurnberg, E. Hanna, A. Kanagaki, Y. Caspi,  D. Raychaudhuri, J. Bechhoefer and the Nikon microscopy center for their help with the experimental part. A.A. was supported by a Junior Fellowship of the Harvard Society of Fellows and the Milton Fund. Work by D.R.N was supported by the National Science Foundation via Grant DMR1005289 and through the Harvard MRSEC via Grant DMR0820484. S.J. was supported by Bauer Fellowship (Harvard University),  Allen Distinguished Investigator Award, Pew Award, and NSF CAREER Award. To whom correspondence may be addressed: arielamir@physics.harvard.edu (A.A.) or suckjoon.jun@gmail.com (S.J.).

\end{acknowledgments}

\newpage

\large
\textbf{Supporting Information}

\normalsize

\setlength{\parindent}{0.25in}
\setcounter{figure}{0}
\section{Microfluidics Setup}
\label{microfluidics}

The central element of the experimental setup consists of a microfluidic device (the ``mother machine"), with dimensions that match well those of the bacteria (width and height of approximately $1$ $\mu$m, matching the bacteria's diameter), see Fig. 1 of the main text. Details of the fabrication of the devices are described elsewhere \cite{jun_}.

In contrast to the work of Ref. \cite{jun_}, where the robustness of the growth was studied and the bacteria were constantly dividing, here we suppress the cell division by regulating proteins involved in the division machinery (SulA for \emph{E. coli} and YlaO for \emph{B. subtilis}), after inserting the cells into the microchannels.  Upon filamentation, a single bacterium will fill the $25$ $\mu$m long channels, and upon further growth part of the bacterium will be exposed to the flow in the main chamber of the microfluidic device, see Fig. 1 of the main text. In most experiments the flow was of LB broth, which was also used to supply the bacteria with nutrients, but we have also done the experiments using minimal and synthetic rich media, see Fig.~5 of the main text. Next, we will characterize the flow and calculate the resulting forces on the bacteria.

\section{Mapping the flow-field}
\label {mapflow}
Throughout the experiment, the Reynolds number is low enough ($\sim 10^{-2}$) such that the flow is laminar. The bacteria reside very close to the surface, though, where the velocity vanishes, and for this reason it is important to accurately characterize the profile of the flow through the device. In the experimental setup, the pressure on the device is controlled. For a given pressure difference, the flow through the device can be analytically expressed as an infinite series~\cite{hashmi_}. Fig.~1a shows the theoretical expectation for the flow through a cross-section of the main channel.

We mapped the flow with spinning-disk confocal microscopy, using fluorescent beads with a diamater of $20$ nm and an exposure time of 2 ms.  We extracted the velocity profile from the lengths of the tracks left by the beads. A comparison of the measurement and the theoretical expectations, with no fitting parameters, is shown in Fig.~1b.

To estimate the drag force on a bacterium (per unit length), we use the exact solution for the viscous drag force (per unit length) on a cylinder residing on a surface \cite{schubert_}:

\be f= 4 \pi \eta r \frac{\partial v}{\partial z} , \ee where $r$ is the bacterium's radius and $\frac{\partial v}{\partial z}$ is the gradient of the velocity profile close to the surface far away from the cell (where the flow velocity is approximately linear in the height above the surface). In our setup, the force is approximately $40$ pN/$\mu$m. In this configuration, the net lift on the cell due the flow can be shown to vanish, but there is a non-vanishing torque per unit length due to the flow \cite{schubert_}:

\be \tau_\text{flow}= 6\pi  \eta r^2 \frac{\partial v}{\partial z}. \ee

This torque leads to a non-vanishing $\sigma_\text{xy}$ component of the stress-tensor. This component, however, is not expected to couple to the cell wall growth and can therefore be neglected. Moreover, the resulting shear stresses obey $\sigma_\text{xy} 2\pi r^2 = \tau_\text{flow}$, and are significantly smaller than the $\sigma_\text{yy}$ stress which we shall later calculate. Similarly, $\sigma_{xx}$ due to the flow is also negligible.

\section{Microscopy and image analysis}
\label{subsubsec:microscopy}

We used a Nikon TI microscope, with fluorescence light provided by the Lumencore Spectra X arc lamp and 100X objective. We used the Nikon Perfect Focus system to maintain the cells in focus during the lengths of each measurement, which typically lasted for 30-60 minutes for each field-of-view. Phase contrast images were taken every 1 second, and fluorescent images with a YFP filter were taken every 20 seconds, in order to minimize the effects of phototoxicity (in Figs. 2a and 3a of the main text we show that exponential growth with the physiological growth rate is achieved for these conditions: slightly less than 30 minutes in Fig.~2A and $\approx 40$ minutes in Fig.~3A). For this reason exposure times were chosen as the minimal ones which provide good signal-to-noise (SNR) ratios. These were typically 100 ms for the phase contrast images and 20-40 ms for the florescent images.

We implemented an image processing algorithm to track the shape of the cell as a function of time, which was optimized for the purposes of our experiment. The phase contrast images were used to initially find the location of the end of the microchannel (not visible in the fluorescent images), manually. For the image processing, only the fluorescent channel images were used, due to their superior SNR (which was typically between 3-5 when comparing the center of the bacterium and the noise level).  The algorithm is described in the following.
\begin{figure*}[t]
\begin{center}
\includegraphics[width= 18 cm]{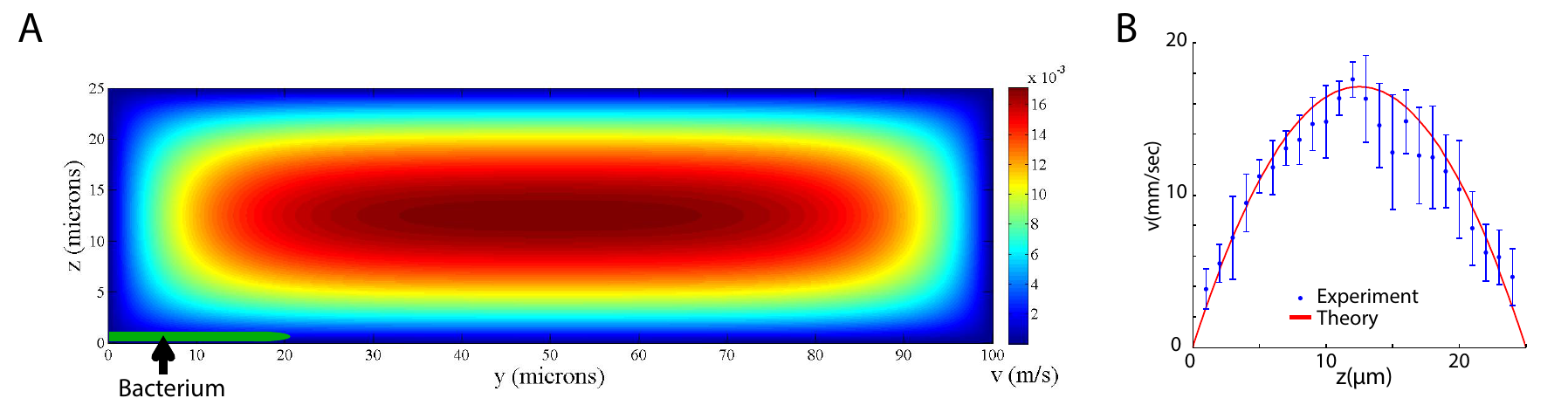}
\caption{A. The results of the analytical solution for the flow through the device. A green bacterial cell protruding about 20 microns into the chamber is superimposed on the flow field at the lower left.
B. Comparison of the measured flow profile, averaged over the width of the device (along the $y$ axis), vs. the theoretical expectation, for a pressure difference of 3 \emph{kPa}.\label{mapflow_fig}}
\end{center}
\end{figure*}

1. Starting at the end of the microfluidic channel, the position along the cell is followed, where at each running step we keep track of the direction of the local tangent to the cell. Initially, the tangent is pointing in the $y$ direction, as shown in point $A$ in Fig.~2.

2. At every step of the algorithm, the \emph{first approximation} to the next tracking point is chosen by taking the current pointer position and advancing it by a small increment (1/15 $\mu$m, corresponding to one pixel) in the direction of the current tangent.

3. The intensity profile is extracted in a line perpendicular to the tangent, see for example the line next to point $B$ in Fig.~2, and the inset showing the corresponding intensity profile. The length of the perpendicular segment is chosen as $1$ $\mu$m, to match the bacteria's diameter - we shall define it as the "search range". The \emph{center-of-mass} of the intensity  in the search range is calculated, in order to minimize the fluctuations, \emph{which is chosen as the new pointer position.}

4. The new tangent can now be chosen as the difference of the updated pointer and the pointer from previous runs, where in order to avoid large local fluctuations due to noise we choose to compare two pointers separated by 20 runs (corresponding to approximately $1$ $\mu$m).

5. To find the tip of the bacteria, we define the noise level as the median of the intensity over the entire field-of-view, and in the previous tracking algorithm we demand that the average intensity over the search range should be greater than a threshold of SNR = 3.

6. When the signal goes below the threshold, implying we have reached the end of the bacterium, we fit a 4th order polynomial to the extracted shape, and used the smooth polynomial in order to calculate the contour length of the bacterium.

\begin{figure}[b]
\begin{center}
\includegraphics[width=0.4 \textwidth]{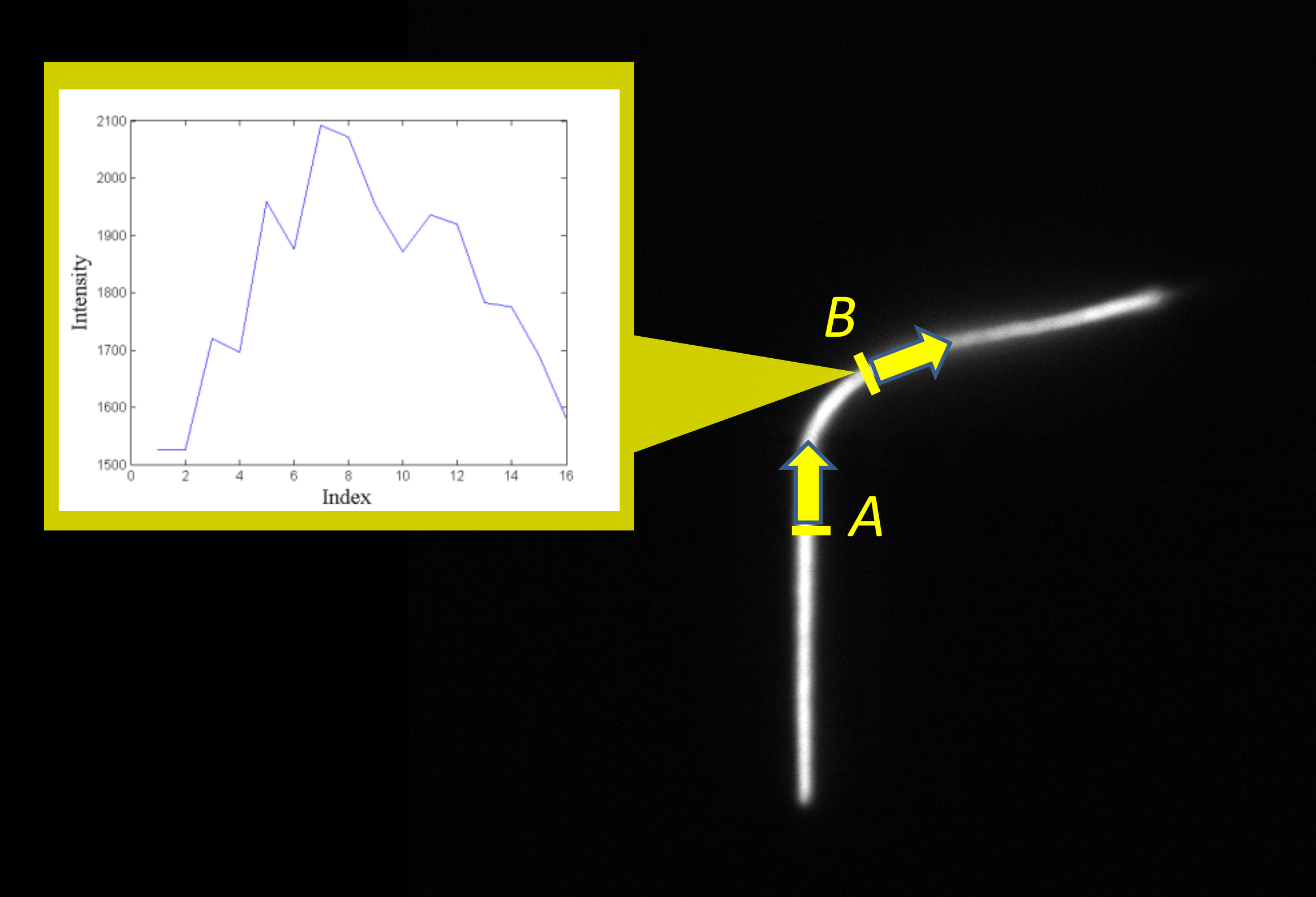}
\caption{\label{algorithm} Schematic illustration of the image processing algorithm used for the analysis. Shown is a raw fluorescent image, adjusted in photoshop using auto-levels for clarity. Starting at point $A$ (the end of the microchannel), at each point the direction of the local tangent is kept (initially in the direction of the microchannel, see the yellow arrow). The intensity profile in a narrow strip perpendicular to the tangent is calculated (see the inset, showing the profile associated with point $B$, reached by the algorithm at a later step), and its center-of-mass is used to update the next point as well as the tangent vector.}
\end{center}
\end{figure}

\section{Theory of the elastic deformations}
\label{theory}

The Young modulus of peptidoglycan was estimated in Refs. \cite{yao_},\cite{shaevitz2_} to be of the order of $Y=3 \cdot 10^7$ N/m$^2$. The thickness of the cell wall, $h$, is of the order of $2-4$ nm. In a similar fashion to the usual derivation of the deformation of an elastic rod, we can calculate the elastic deformation of a hollow rod of thickness $h \ll r$, subject to a constant force per unit length perpendicular to its axis of symmetry. This is equivalent to the problem of a rod being deformed under its own weight, with one fixed end and one free end, see Ref. \cite{LL_elasticity_} p. 81. We outline the details of the calculation, since part of the reasoning will also be relevant to the discussion of the \emph{irreversible} deformations in section \ref{plastic_sect}. A key ingredient is the (geometric) connection between the local radius-of-curvature and the local strains. As illustrated in Fig.~3, for a given radius-of-curvature $R(y)$ at a given point $y$ along the cell, the strain on a smaller region of the outer part of the cell wall will be given by:

\begin{figure}[b]
\begin{center}
\includegraphics[width=0.5 \textwidth]{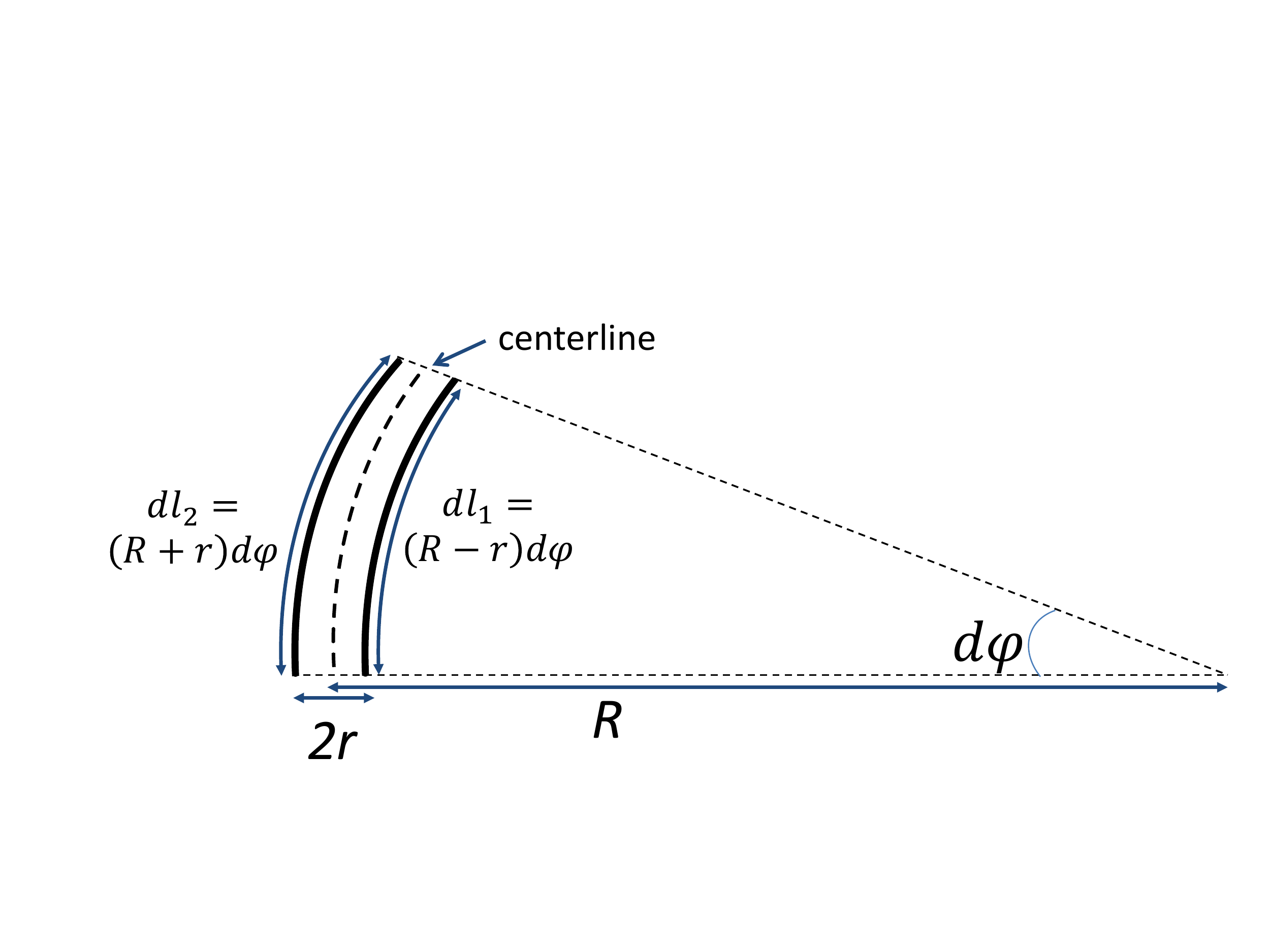}
\caption{\label{bent} A small section of a bent cell. }
\end{center}
\end{figure}

\be \frac{\delta l}{dl}=\frac{dl_2-dl_1}{R d\varphi} \approx 2r/R, \label{curvature} \ee
where $dl=R d\varphi$ is the arc traversed by the centerline.

 Here, we assume that the strains vanish for a straight cell, an assumption that will \emph{not} hold when we will consider asymmetric insertions in section \ref{plastic_sect}. We also assumed that the strains are small, which does \emph{not} imply that the deflections are small - for a cell with a large aspect ratio, one can have small strains throughout the cell leading to an overall large angular deflections. Typical deflections $\delta(y)$ calculated below are in the range $r \ll \delta(y) \ll l$, where $l$ is the protruding length of the bacterium. In the regime of linear elasticity, the strains are proportional to the stresses. The surrounding viscous flow creates primarily a $\sigma_\text{yy}$ component of the stress tensor within the cell wall (see Fig. 1 of the main text for the definition of the coordinate system), which can be found from the condition that the torques created by them must compensate for the torque created by the external forces.
We proceed to express the extra stress at a point with coordinate $y$ along the centerline of the cell's long axis and azimuthal angle $\theta$, in terms of the flow parameters (see Fig. 6 for the definition of $\theta$). Using the linear relation of stress and strain in a hollow rod \cite{LL_elasticity_}, we expect that for small deformations:
\be \sigma_\text{yy}(\theta,y) \approx Y r \sin(\theta)/R(y), \label{stress_R}\ee where $R(y)$ is the local radius-of-curvature (not to be confused with the radius of the bacteria, $r$), and $Y$ is the Young modulus of the cell wall. Upon balancing the torque induced by this stress with the external torque at position $y$, $\tau(y)$, we find that:

\be \tau(y) = \frac{Y I}{R(y)}, \label {torque} \ee
where $I$ is the `moment-of-inertia' of a cross-section, which for our case is found to be $I=\pi r^3 h$ (this result can also be obtained by differentiating the formula for a solid cylinder, $I_s=\frac{\pi}{4}r^4$). Note that the moment-of-inertia is a mathematical analogy with the theory of rigid bodies, a notation commonly used in the theory of rod elasticity. $YI$ is known as the flexural rigidity. For a constant force per unit length $f$ we find that the stresses within the cell wall are:

\be \sigma_\text{yy}(\theta,y) = \frac{f \sin(\theta)(y-l)^2}{2\pi r^2 h}, \label{stress}\ee
This result should be multiplied by the thickness $h$ if one is interested in the two-dimensional stress. Indeed, one can check directly that the torques arising from these stresses at a cross-section at point $y$ are equal to the total torque due to the force $f$ integrated along the segment $[y,l]$. Upon combining Eq. (\ref{stress}) with Eq. (\ref{stress_R}), the resulting curvature is given by:
\be 1/R(y)\approx \frac{f (y-l)^2}{2YI} \label {torques_elastic}. \ee

We note from differential geometry that the radius-of-curvature can be expressed in terms of the horizontal deflection of the bacterium $\delta(y)$, as:

\be \frac{1}{R}=\frac{\delta''}{(1+\delta'^2)^{3/2}}. \label{full_elastic}\ee
Upon neglecting the $\delta'$ term in the denominator, valid for weak deflections, and focusing on the case of a constant force along the cell,  we obtain a linear equation that can be readily integrated to give the resulting deflection of a point $y$ along the cylinder. Under the assumption of a vanishing derivative at the end of the microchannel, we find that:
\be \delta(y) =\frac{f y^2(y^2-4ly+6l^2)}{24 Y I} ,\label{z_} \ee

Thus, the maximal deflection $\delta_\text{max}=\delta(y=l)$ is given by:
\be \frac{\delta_\text{max}}{l}=\frac{f l^3}{8 Y I} . \label{elastic_bend} \ee%

Thus, the horizontal deformation of the tip $\delta(l)$ is approximately proportional to $l^4$, where $l$ is the length of the cell exposed to the force. Based on the data in Fig. 2b of the main text, we were able to calculate the stiffness of the \emph{E. coli} cells, and extract a flexural rigidity of $4-6\cdot 10^{-20}$ N $\cdot$ m$^2$. This result is similar to that obtained by very different methods such as AFM~\cite{yao_} and recent work in which the cells were transiently deformed by optical traps~\cite{wang_}. It should be pointed out that even though the perpendicular deflections are significant, the local strains and stresses are in fact small. Thus, it is justified to assume that we are in the regime of \emph{linear} elasticity. Due to the large aspect ratio of the filamentous bacteria, whose length $l$ is much larger than their radius $r$, the deflections are enhanced by a factor of $l/r$.

Recently, bending of elastic fibers in a flow was studied, experimentally and theoretically, albeit in a different regime of parameters where the flow is highly confined and the fiber blocks most of it \cite{stone_}.

\section{The effect of the bacterium/microchannel mismatch on the elastic measurements}
\label{elastic_corrections}

The previous analysis of the elastic deformations assumed a boundary condition with the cell fixed at the end of the microchannel, i.e., that the tangent to the cell centerline would be parallel to the microchannel there. However, looking at a typical cell (see for example Fig.~5 shows that there can be a finite derivative $\delta'(y=0)=a $ there, due to the small mismatch between the cell diameter and the channel diameter at the end of the microchannel, which we  denote by $\Delta$.

We will find $a$ by analyzing the forces and torques, show that the assumption of a cell fixed at the microchannel end is an excellent one for the longer cells, and account for the deviations for the shorter cells.

The torque balance for the part of the cell outside the microchannel implies that $y''$ is still given by Eq. (\ref{torques_elastic}), and thus the shape of the cell is given by:

\be \delta(y) =\frac{f y^2(y^2-4ly+6l^2)}{24 Y I}+a y ,\label{z} \ee
where $a$ is yet to be determined. For the part of the cell inside the microchannel, external forces $F_1$ and $F_2$ are only applied at two points, shown in Fig.~5. The derivative at the point where $F_2$ is applied vanishes (since the cell is tangent to the microchannel there), so the position of the centerline of the cell is given by the usual solution of a beam bent by an external force \cite{LL_elasticity_, wang_}:
\be \delta(y) = \frac{F_2}{YI} \left(\frac{y_0(y_0+y)^2}{2}-\frac{(y_0+y)^3}{6}\right), -y_0\leq y \leq 0\label{delta} \ee
where $y=0$ corresponds to the end of the microchannel.
We therefore have:
\be \Delta = \frac{F_2}{3YI}y_0^3. \label{eq1_elastic} \ee
However, both $F_2$ and $y_0$ are unknown. We fix these quantities by considering the torques around the end of the microchannel:
\be F_2 y_0 = \tau = \int_0^l f l' dl'=f l^2/2. \label{eq2_elastic} \ee
Upon eliminating $F_2$ from Eqs. (\ref{eq1_elastic}) and (\ref{eq2_elastic}), we find that:
\be \frac{1}{3YI}y_0^2 = \frac{\Delta}{\tau}. \ee
Hence:
\be y_0 = \sqrt{\frac{3 YI \Delta}{\tau}} , F_2=y_0/\tau=\frac{2 y_0}{f l^2}. \ee
Since the derivative at the end of the microchannel is continuous, we have:
\be a =  \sqrt{\frac{3 \Delta \tau}{4 YI }}=\sqrt{\frac{3 \Delta f l^2}{8 YI }}. \ee

For the case of a perfect match between the bacterium and the microchannel diameter ($\Delta =0$), the maximal deflection was given by:
\be \delta_\text{max} =  \frac{f l^4}{8 Y I}. \ee
The corrected maximal deflection will now be:
\be \delta_\text{max} =  \frac{f l^4}{8 Y I} + l^2 \sqrt{\frac{3 f \Delta}{8 Y I}}. \label{correctdelta} \ee
The correction is negligible for long enough cells, $l \geq (24 Y I /f)^{1/4} $, but becomes significant for short cells.
It is also possible to take into account a \emph{non-uniform} force per unit length outside the channel: in this case, the force per unit length $f$ and torque $\tau$ must be recalculated as a function of position on the cell.   The deflection can then be calculated by numerical integration. Fig.~2b in the main text uses this more general procedure when modeling the elastic deformations:

1. We use the full solution for the flow profile to calculate the force per unit length at every point along the cell, and the resulting torques at every point.

2. We solve the Elastica problem for this torque distribution, and find the resulting deflection by integrating the equations numerically, including the non-linear terms due to the denominator of Eq. (\ref{full_elastic}).

For stiff, long cells using Eq. (\ref{correctdelta}) with a constant force per unit length $f \approx 40$ pN/$\mu$m would be an excellent approximation, since the flow would be approximately constant and the deformations of stiff cells would be still small such that the non-linear contributions are insignificant. This is the case for the data shown in Fig.~4 below.

\section{Simple and non-invasive measurements of Young's modulus}
Transient bending by pulse-like forces provide simple, non-invasive means to directly measure the elastic properties of living bacterial cells. Fig.~4 shows the maximum deflection $\delta_\text{max}$ vs. the length ($l$) of the cells exposed to the force. The solid red line is the fit using a full theoretical flow profile in Fig.~1, and we obtained the flexural rigidity $Y I \approx 2.4 \times 10^5 \text{pN} \cdot \mu \text{m}^2$. This implies $Y \approx 20.4$ MPa for a cell radius $r = 0.5$ $\mu$m and cell-wall thickness $h=30$ nm, in good agreement with past measurements which found Y = 10 -- 50 MPa \cite{Thwaites1989_, Thwaites1991_, Mendelson2000_}.

We also used Eq.~\ref{correctdelta} to fit both the average force per unit length, $f$, and the flexural rigidity ($\Delta$ can be measured directly from the images). We obtain a fit that is virtually identical to the full-flow result (Fig.~4). For $\Delta = 0.5$ $\mu$m, we find $f \approx 41.2$ pN/$\mu$m and $YI \approx 2.1 \times 10^{-19}$ N$\cdot$m$^2$. With the same assumptions from above we find $Y \approx 19.9$ MPa, consistent with the results using the full flow profile.  This suggests that our flow-based scheme of imparting force may provide a simple and robust measurement of the Young's modulus of various bacteria.

\begin{figure}[b]
\begin{center}
\includegraphics[width= .45 \textwidth]{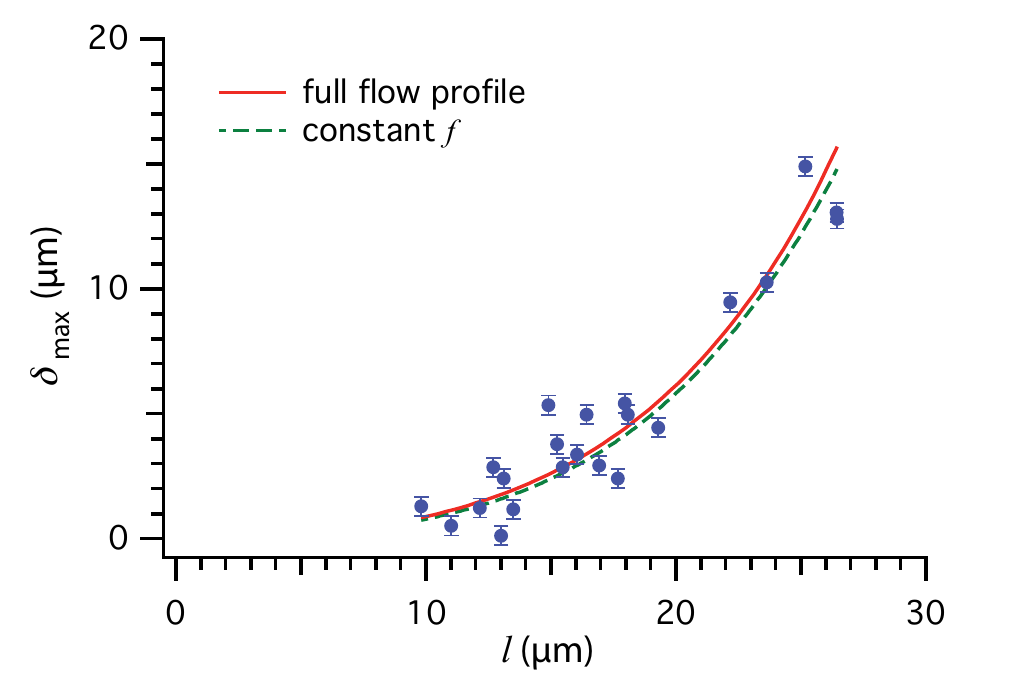}
\caption{\label{BsubElastic} Elastic measurements of \emph{B. subtilis}.  Cells were filamented in the absence of flow and perturbed with brief $\sim 15$ s pulses of media flow.  The resulting maximum displacements $\delta_\text{max}$ are plotted as a function of the exposed cell length $l$ for the elastic deformations.  The red line shows the elastic theory results using the full nonlinear flow profile (in analogy to Fig.~2 of the main text for \emph{E. coli}), yielding $YI\approx 2.4 \times 10^{-19}$ N$\cdot$m$^2$.  The green dashed line is a fit to equation \ref{correctdelta}; holding $\Delta = 0.5$ $\mu$m, we find $f \approx 41.2$ pN/$\mu$m and $YI \approx 2.1 \times 10^{-19}$ N$\cdot$m$^2$.}
\end{center}
\end{figure}

\begin{figure}[b]
\begin{center}
\includegraphics[width=0.3 \textwidth]{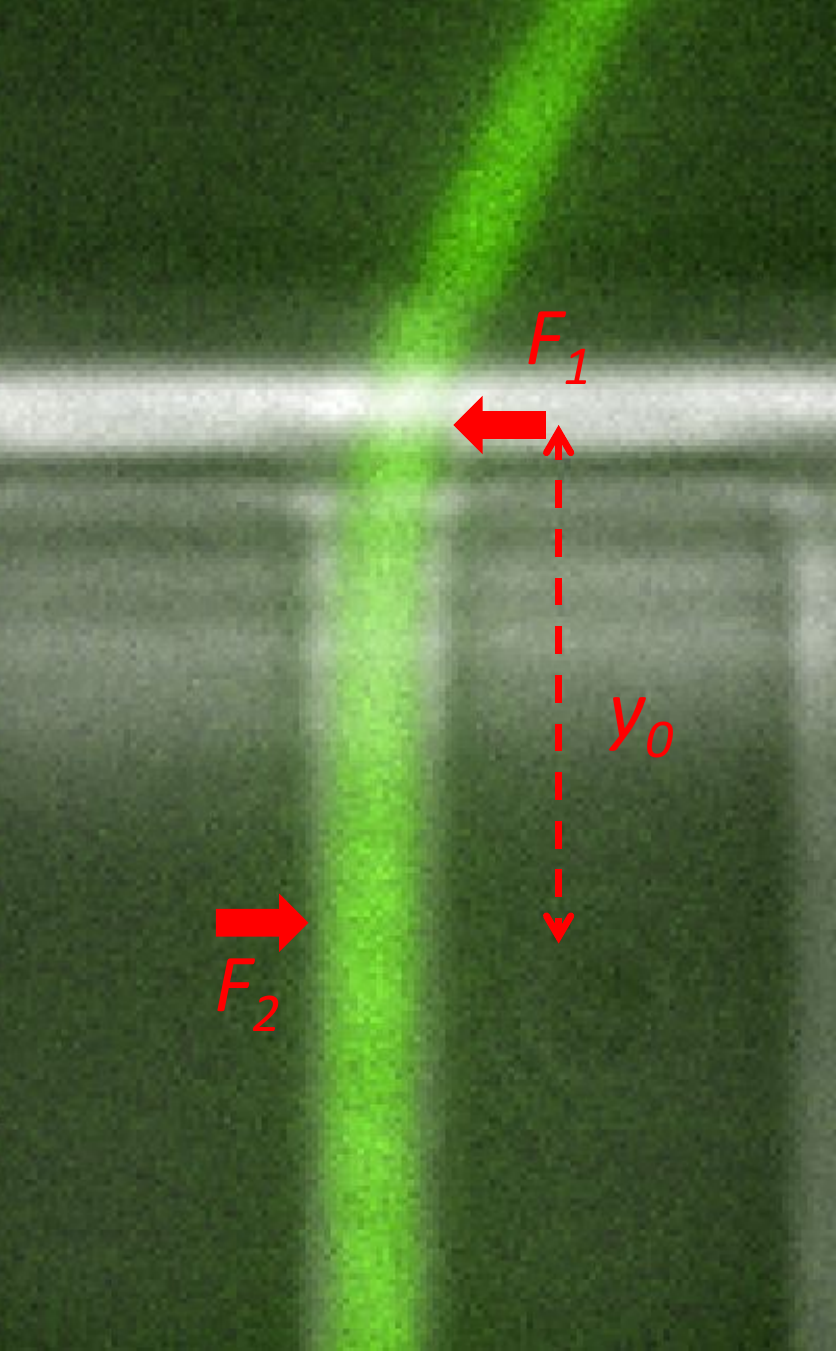}
\caption{\label{channel_analysis} Forces on a bacterium confined in a channel.}
\end{center}
\end{figure}

\section {The lack of torque due to turgor pressure}
The turgor pressure creates no torques on the cell, and for this reason need not be taken into account throughput the analysis. To show this, consider an integral over the closed red cross-sectional contour surface sketched in Fig.~6. The force and torque due to these forces must vanish, since the material inside is in mechanical equilibrium. In the analysis of cell wall deformations, one has to find the torque exerted on the cell wall by external forces. The contribution to this torque by the turgor pressure around point $A$, for example, is exactly that associated with the above contour, except for the missing additional torque associated with the flat surface inside the cell (denoted by the dashed line in the figure). The forces there, however, give rise to a negligible torque, since they have no lever arm. We conclude that the torque due to turgor pressure around any point along the cell would vanish. The only effect of the pressure is thus to create large stresses $\sigma_{xx}$ and $\sigma_\text{yy}$ in the cell wall.

\begin{figure}[b]
\begin{center}
\includegraphics[width=0.2 \textwidth]{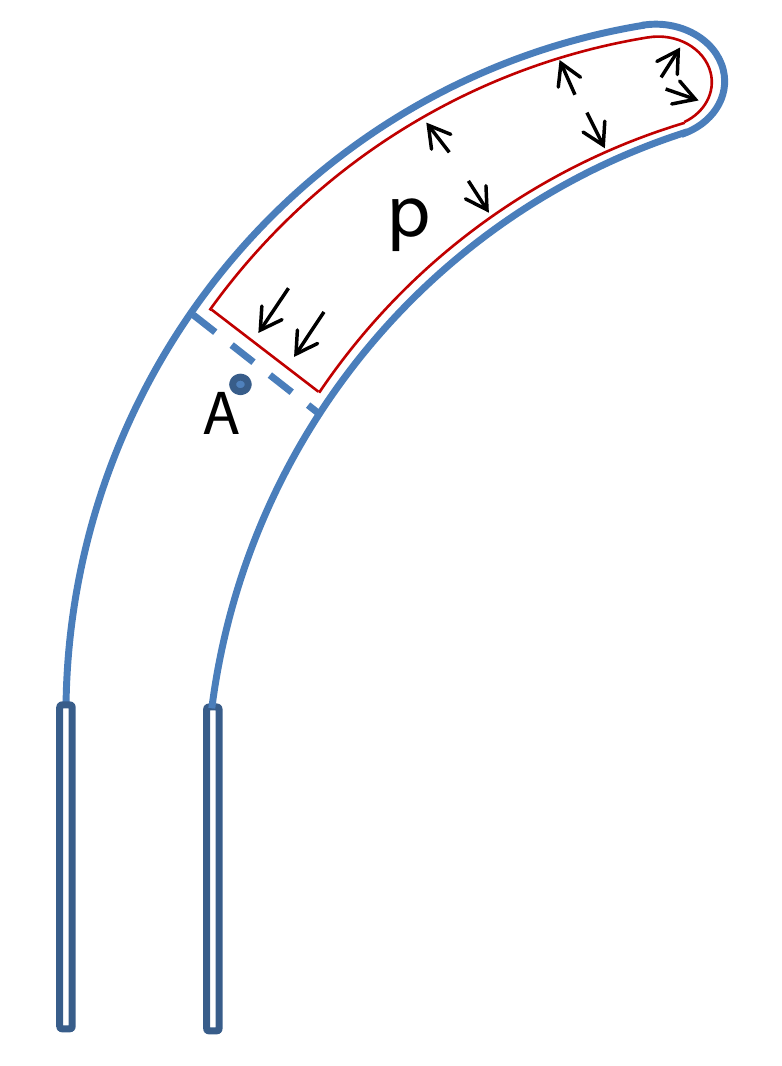}
\caption{\label{contour} Turgor pressure exerting forces on a closed contour just inside the cell wall of a protruding bacterium.}
\end{center}
\end{figure}

\section {Theory of the plastic deformations}
\label{plastic_sect}

 We will now estimate the rate of bending due to the asymmetric synthesis of cell wall as a response to an external stress (due to the viscous flow). This asymmetry gives rise to an additional bending mechanism, which is irreversible, since it involves remodeling of the peptidoglycan mesh. We will assume that we are still in a regime where the bending is relatively small, so we can use the previous results for the flow around a straight cylinder, and for the forces on a slightly deformed rod. For simplicity of the analysis, let us start by assuming that the Young modulus is large enough such that elastic bending is negligible; we shall see that even for $Y \rightarrow \infty$ the plastic bending is finite. In the next section we will explain how the calculation can be extended to take both elastic and irreversible plastic deformations into account.
\begin{figure}[t]
\begin{center}
\includegraphics[width= 8 cm]{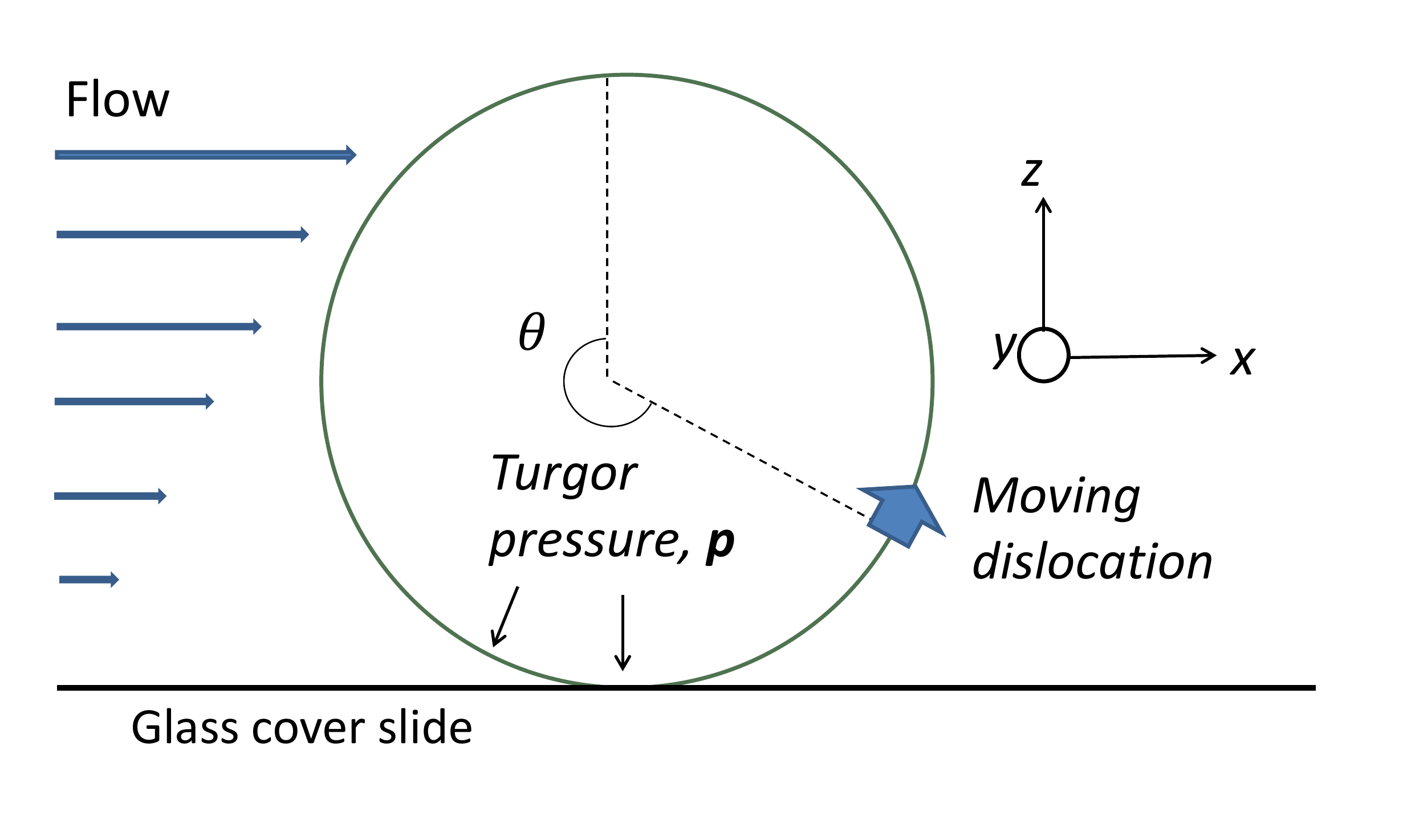}
\caption{Cross-section of a growing bacterium.}
\label{cross_sect}
\end{center}
\end{figure}

Consider a cross-section of the bacterium, at a distance $y$ away from the end of the microchannel, as illustrated in Fig.~7. The \emph{two-dimensional} stress due to the turgor pressure $p$ is independent of $\theta$, and given by $\sigma^0_\text{yy}= pr/2$. Eq. (\ref{stress}) shows that the additional stress $\sigma_\text{yy}$ due to the viscous drag is positive for $0<\theta <\pi$, corresponding to the stretched part of the rod, and negative otherwise. The maximal relative change in the stress occurs at the points $\theta=\pm \pi/2$, at $y=0$ (corresponding to the end of the microchannel) and equals:
\be \Delta \sigma/\sigma_\text{turgor} = \pm \frac{f l^2}{\pi p r^3}. \ee
Upon inserting the numbers for a relatively long filamentous bacterium of length 10 $\mu$m, we find a ratio of order unity.  
Thus, the flow can induce significant changes in the stress, for long cells.

We shall now invoke the formalism of dislocation-mediated growth developed in Ref. \cite{amir_nelson_pnas_}, which hinges on the fact that new material is inserted circumferentially via the rotating strand-extension centres. The $\sigma_\text{yy}$ stress induces a Peach-Koehler force on an edge dislocation with a Burgers vector in the $y$ direction, acting in a tangential direction (in the $xz$ plane), $F=b\sigma_\text{yy}$, as discussed in the main text. This force acts in addition to the always positive stress $\sigma^0_\text{yy}$ due to turgor pressure.
Tensile stresses enhance growth while compressive stresses inhibit it. Under the assumption that the additional stresses are small compared to that due to turgor pressure, we can Taylor expand the dependence of local growth rate on stress, and find:

\be \Delta g /g = f[\sigma_\text{yy} / \sigma^0_\text{yy}] \approx \alpha \frac{\sigma_\text{yy}}{pR/2}, \label{taylor} \ee
where the coefficient $\alpha$ multiplying the excess stress $\sigma_\text{yy}$ due to bending is expected to be of order unity.
Upon inserting the additional stress when a force $f$ per unit length is exerted on the cell, we find that:

\be \Delta g /g \approx \alpha \frac{f l^2}{2\pi p r^3 }, \label {assymetry} \ee
(where we take $y=l/2$).

This differential growth, accumulated over time, will turn out to give a non-negligible plastic deflection even when $\Delta g / g \ll 1$, due to the large aspect ratio of the filamentous bacterium. The bending of the cell will be such that the additional strains corresponding to the bending will compensate those due to the differential growth: locally, the ratio of arcs $dl_2/dl_1$ in Fig.~3 should correspond to the relative amount of material incorporated into the cell wall. Let us denote the \emph{integrated} differential growth arising Eq. (\ref{assymetry}) by $\Delta l$.  The radius-of-curvature $R$ resulting from the differential grows (see Fig. 3) obeys:

\be \Delta l / dl \approx  (dl_2-dl_1)/dl_1= 2r/R, \label {growth_curvature} \ee
leading to a curvature $\kappa \equiv 1/R = \frac{1}{2r} \Delta l / dl$. The angle $d\varphi$ resulting from this differential growth obeys $R d\varphi = dl$, hence:
\be d\varphi = \frac{dl}{2r} \Delta l / dl .\ee

Indeed, we find that the relative growth asymmetry of $\Delta l / dl$ is enhanced by the geometrical factor of $\frac{dl}{2r}$. For a long cell with a constant differential growth, this factor would correspond to half the aspect ratio. Note, however, that this argument gives the \emph{local} curvature due to differential growth, which in our experiments is non-constant along the length of the cell. These results can be also obtained via differential geometry, by considering the Gaussian curvature associated with the non-uniform metric due to differential growth.	

We thus expect that as the bacterium grows to length $l$, the angle associated with the plastic bending (as long as it is small) should scale as:

\be \phi \propto \frac{\Delta l}{l}\frac{l}{2r} \sim \frac{f l^3}{p r^4 } . \label{scaling_plastic}\ee
Upon comparing this angle to that of the elastic deformations (Eq.~\ref{elastic_bend}) , we find that both elastic and plastic effects have the same scaling with $l$, allowing for a quantification of their relative importance in terms of a single dimensionless parameter:
\be \chi \equiv \frac{p r}{Y h} . \label {plastic_vs_elastic} \ee For large enough $Y$, the plastic effect dominates over the elastic one.
With estimated values for \emph{E. coli}, we find that $\chi$ is of order unity, implying that elastic and plastic effects are of the same order of magnitude. Thus, one has to solve the coupled problem, without separating the discussion into an elastic and plastic part, as we did so far. For bacteria with stiffer cell walls, we expect the plastic effect to dominate over the elastic one.

\begin{table}
\centering
\caption{Summary of parameters for calculating $\chi$ for \emph{E. coli} and \emph{B. subtilis}.  $^*$ refers to the present work. }\label{ChiTable}
\begin{tabular}{ l | c c | c c }
  & \emph{E. coli} & Ref. & \emph{B. subtilis} & Ref.\\
 \hline
  $p$ (atm) & 0.3; 2--3 & \cite{shaevitz2_}; \cite{Koch1990_} & 20 &  \cite{whatmore_}\\
  $r$ ($\mu$m) & 0.3--0.5 & \cite{Zaritsky1978_} & 0.4 & \cite{Maass2011_}\\
  $Y$ (MPa) & 30 & \cite{yao_, shaevitz2_},$^*$ & 10--30; 50; 20 & \cite{Thwaites1989_, Thwaites1991_}; \cite{Mendelson2000_};  $^*$\\
  $h$ (nm) & 1.5--6.5 & \cite{Vollmer2008_} & 30 & \cite{Beeby2013_}\\
  \hline
\end{tabular}

\end{table}

Finally, we note that when going from Eq. (\ref{assymetry}), giving the instantanous relative growth, to the integrated asymmetry $\Delta l$, one should also take into account the effect of a finite processivity, which in the absence of differential growth would lead to an exponential decay of the curvature \cite{jacobs_}. The accumulated \emph{relative} differential growth $\rho \equiv \Delta l / dl$ obeys:
f
\be \frac{d \rho}{dt}= -\lambda \rho + \alpha \frac{\sigma_\text{yy}}{pR}\lambda, \label{integrate} \ee
where $\lambda$ is the growth rate (i.e., the doubling time equals $t_d=\log(2)/\lambda$), the first term corresponds to the ``dilution" effect due to the finite procreativity (taken as infinite in the above equation, for simplicity), and the second term corresponds to Eq. (\ref{taylor}).


\section{Straightening after snapback}

The straightening described in the main text (see also Fig.~3 of the main text and Supplementary Movie S2) is in sharp contrast to the results of Ref. \cite{whitesides_}, where curved, arc-shaped cells taken out of microfluidic channels do not seem to straighten as they grow, but rather, maintain a self-similar shape with a slowly decreasing radius-of-curvature as they become larger. In fact, the latter is precisely what is expected from circumferential insertions with long processivity \cite{mukhopadhyay_, sean_}: circumferential insertions would ``dilute" the assymtery due to differential growth and thus lower the curvature, yet since the cell grows longer the shape can be shown to be self-similar. Taking into account a finite processivity of insertions would make the decay of the curvature slower \cite{jacobs_}. Therefore, if processive insertions alone are responsible for straightening, during continued growth the tip should move towards the bending direction ($\rightarrow$). In contrast, the cells after the partial snap back, however, actually move towards ($\leftarrow$) the channel the bent cell is embedded in (Fig.~3b of the main text).

Note that due to the turgor pressure, the stress on the cell wall during the straightening process is non-uniform, and one should also take the coupling of the stress distribution to the growth. However, when a cell is deformed, the stress component along the bacterium's long axis ($\sigma_\text{yy}$) remains approximately constant, while it is the circumferential component of the stress tensor that is modified, in a manner proportional to the curvature of the bent cell \cite{shells_, jacobs_}. There are no indications that this stress component is coupled to the growth process. We therefore conclude from our experiments that another mechanism is at play, which makes the cell straightening significantly faster in the absence of mechanical forces. One possibility is a curvature-sensing mechanism that has been discussed in the context of protein localization in bacteria~\cite{losick_}. A second mechanism is the appearance of residual stresses due to the differential growth in the first stage of the experiment: the combination of turgor pressure and the differential growth can lead to residual stresses in the cell wall, which remain even after the flow is turned off, and are in general non-uniform. We leave the study of this intriguing phenomenon to future work.

\section{Plastic deformation relies on cell growth}
In the main text we demonstrate that plastic deformation of growing cells results from the application of flow-generated hydrodynamic force over extended periods of time.  We further demonstrate that, in the time after the flow is stopped, the cells straighten as they grow.  We attribute this behavior to a dislocation-mediated growth mechanism in which anisotropic stresses give rise to varying cell-wall synthesis rates which leads to plastic deformation.
\begin{figure*}[t]
\begin{center}
\includegraphics[width=\textwidth]{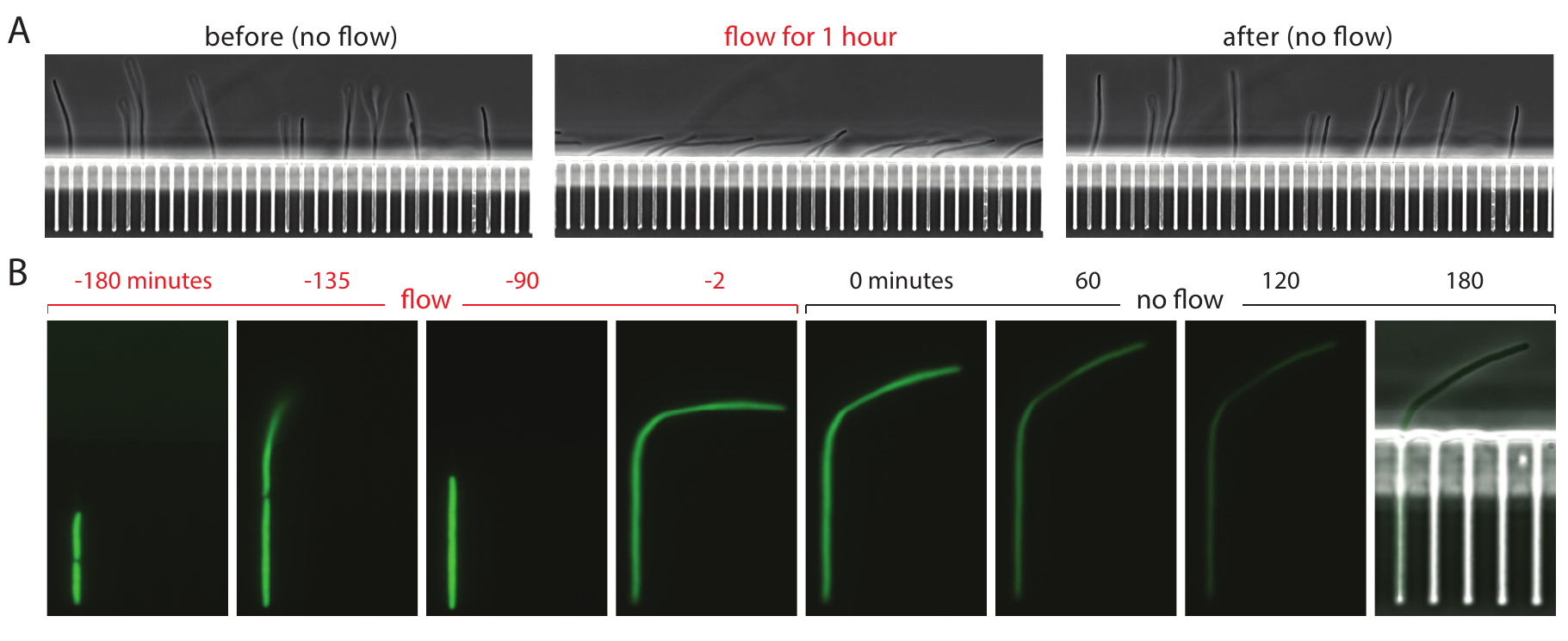}
\caption{Controls: Non-growing cells do not plastically deform.  A.~Montage showing that non-growing cells do not deform plastically under flow.  \emph{E. coli} cells are filamented in the absence of flow until they protrude from the channels by $\sim$ 25 $\mu$m.   At that point, we begin excessive fluorescent illumination and the cells stop growing within 20 minutes.  We then initiate flow, bending the cells and maintain constant flow for 1 hour.  After this period we stop the flow and the cells snap back and fully recover their shape. Thus, non-growing cells do not deform plastically and the Stokes force during our experiments does not mechanically damage the cells. B.~Montage showing that non-growing cells do not straighten.  Filamentation is induced in an \emph{E. coli} cell and it grows in the presence of flow until t = 0.   Beginning at t = -24 minutes, the cell is exposed to excessive fluorescent illumination which continues for the remainder of the experiment, inhibiting growth.  At t = 0 snapback was performed by arresting the flow.  From t = 60 to 180 minutes, the cell shows no appreciable growth, intracellular GFP is photobleaching, and the cell does not straighten.  In the final frame, a phase-contrast image is overlaid, showing the channels and the cell outline.}
\label{nongrowing_fig}
\end{center}
\end{figure*}

   A remaining possibility, however, is that this plastic deformation is disconnected from growth and the cell wall passively reorganizes (deforms) in response to the anisotropic stresses.  To eliminate this mechanism we ran two separate controls.  First, we show that non-growing cells do not plastically deform after extended application of flow (Fig.~9A).

  Second, we demonstrate that cells grown in (and plastically deformed by) flow do not straighten over time (Fig.~9B).  Taken together and in contrast to Fig.~3 of the main text, these experiments indicate that the plastic deformation quantified in this paper is connected to growth of the cell, consistent with a dislocation-mediated growth mechanism.

\section{Shape of a plastically deformed cell}
\label{shape_complete}

We have shown above that filamentous bacteria will have significant elastic as well as irreversible bending at a length $ 10-20$ $\mu$m. In this section we explain how both of the effects can be quantitatively combined to calculate the shape of the cell at a given point.

For the elastic effect, the radius-of-curvature at a given point is proportional to the strain at that point. For the irreversible effect, it is proportional to the non-uniformity in insertion rates on the two sides of the bacterium (relative to the flow direction). Since both effects turn out to be of the same order-of-magnitude for the experimentally relevant case, we have to consider both contributions simultaneously . Upon repeating the previous arguments, we find that:

\be \frac{1}{R(y)}=\frac{1}{R_e(y)}+ \frac{1}{R_i(y)} ,\ee i.e., the curvature is the sum of the elastic (reverible) and growth-induced (irreversible) contributions.
The elastic contribution is given by Eq. (\ref{torque}):

\be \frac{1}{R_e(y)}=\frac{\tau(y)}{Y I} ,\label {c_e}\ee
where $\tau(y)$ is the torque due to the flow, while the growth-related contribution is given by Eq. (\ref{growth_curvature}):

\be \frac{1}{R_i(y)}=\frac{1}{2r}\frac{\Delta l(y)}{dl} . \label {c_i}\ee

\begin{figure}[b]
\includegraphics[width=8 cm]{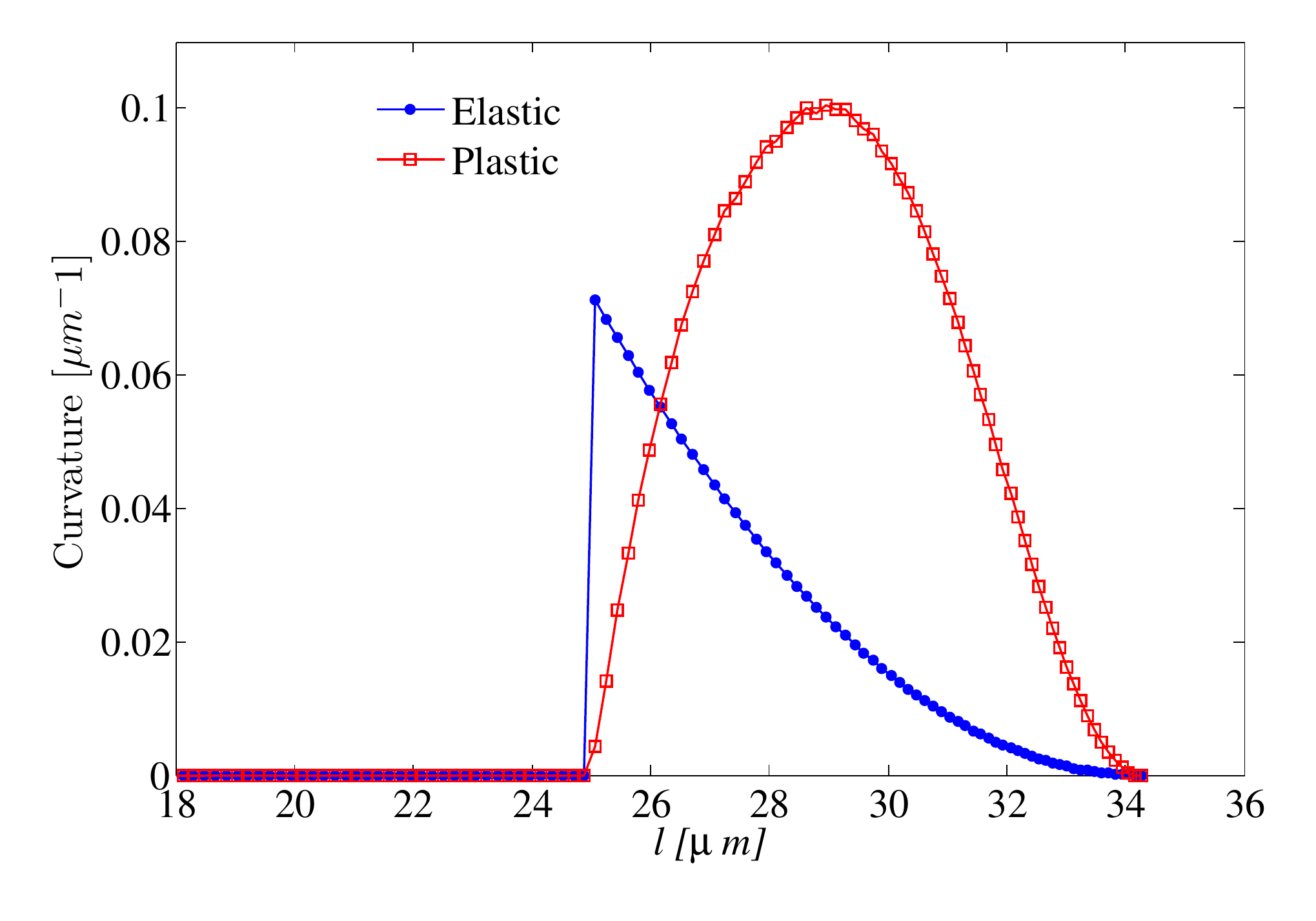}
\caption{Contribution of the elastic (reversible) and plastic (irreversible) effects to the cell curvature, as found by numerically solving the coupled equations for differential growth and for the elastic response, with $\alpha=2.5$ in Eq. (\ref{taylor}). The elastic effect is maximal just outside the microfluidic channel, where the stresses are maximal, and vanishes at the stress-free tip of the bacterium. The plastic effect, on the other hand, relies on accumulation of differential growth over time, and therefore vanishes both at the stress-free tip and at the end of the microfluidic channel, which contains "fresh" cell wall with no asymmetry, having just come out of the microfluidic channel.}
\label{curvature_figure}
\end{figure}

We can now quantitatively analyze the experimental scenario: at every point in time we calculate the resulting torque $\tau$ due to the flow.
This torque gives rise to an elastic curvature (as quantified in Eq. (\ref{c_e})), which is an instantaneous effect. It will also give rise to a non-uniform insertion rate, which, \emph{integrated over time} (see Eq. (\ref{integrate})), will give rise to the irreversible component of the curvature (as quantified in Eq. (\ref{c_i})). Note that a given point on the cell elongates exponentially in time, and that when calculating the accumulated asymmetry  $\frac{\Delta l}{dl}$ at a given point we have to follow the path of the point over time, and integrate the contribution of the stresses at different points in space and time, associated with the ``history" of that part of the cell wall. For example, the part of the bacterium which is close to the end of the side channels had no stress on it while it was in the channel, and for this reason accumulated no asymmetry: it will have no radius-of-curvature associated with the irreversible effect. It will have, however, the \emph{maximal} radius-of-curvature associated with the elastic contribution, since the torque at that point is maximal.  Fig.~8 shows the relative contributions of both effects, for realistic experimental parameters.

%

\end{article}

\end{document}